\documentclass[preprint,5p,twocolumn]{elsarticle}
\usepackage{graphicx}
\usepackage{amssymb}
\usepackage{amsmath}
\usepackage{xcolor}
\usepackage{subfig}
\usepackage{lineno}
\usepackage{hyperref}

\usepackage{xspace} % fixes space issues after command
\newcommand{\alal}{$\alpha$-$\alpha$\xspace}
\newcommand{\bega}{$\beta$/$\gamma$\xspace}

\newcommand{\mevee}{MeV$_{ee}$\xspace}
\newcommand{\dt}{$\Delta T$\xspace}
\newcommand{\capt}{shown for crystal C6 as an example\xspace}

\journal{Astroparticle Physics}

\begin{document}

\begin{frontmatter}



%% Title
\title{Alpha backgrounds in NaI(Tl) crystals of COSINE-100}

%% Authors

\cortext[corr]{Corresponding authors}

%\author[ibs,ust]{H.~Lee\corref{corr}} \ead{gksmrf222333@naver.com}
%\author[ibs,skku]{G.~H.~Yu\corref{corr}} \ead{tksxk752@naver.com}
%\author[wipac]{M.~Kauer\corref{corr}} \ead{mkauer@physics.wisc.edu}
%\author[ibs]{E.~J.~Jeon\corref{corr}} \ead{ejjeon@ibs.re.kr}

\author[yale]{G.~Adhikari}
\author[usp]{N.~Carlin}
\author[usp]{D.~F.~F.~S. Cavalcante}
\author[ibs]{J.~Y.~Cho}
\author[ibs,snu]{J.~J.~Choi}
\author[snu]{S.~Choi}
\author[shef]{A.~C.~Ezeribe}
\author[usp]{L.~E.~Fran{\c c}a}
\author[cau]{C.~Ha}
\author[ewu,cens,ust]{I.~S.~Hahn}
\author[yale]{S.~J.~Hollick}
\author[ibs,ust]{E.~J.~Jeon\corref{corr}} \ead{ejjeon@ibs.re.kr}
%\author[yale]{J.~H.~Jo}
\author[snu]{H.~W.~Joo}
\author[ibs]{W.~G.~Kang}
\author[wipac]{M.~Kauer\corref{corr}} \ead{mkauer@physics.wisc.edu}
\author[ibs]{B.~H.~Kim}
\author[knu]{H.~J.~Kim}
\author[cau]{J.~Kim}
\author[ibs]{K.~W.~Kim}
\author[ibs]{S.~H.~Kim}
\author[snu]{S.~K.~Kim}
\author[ibs]{S.~W.~Kim}
\author[ibs,ust]{W.~K.~Kim}
\author[ibs,ust,su]{Y.~D.~Kim}
\author[ibs,kri,ust]{Y.~H.~Kim}
\author[ibs]{Y.~J.~Ko}
\author[knu]{D.~H.~Lee}
\author[ibs]{E.~K.~Lee}
\author[ibs,ust]{H.~Lee\corref{corr}} \ead{gksmrf222333@naver.com}
\author[ibs,ust]{H.~S.~Lee}
\author[cens]{H.~Y.~Lee}
\author[ibs]{I.~S.~Lee}
\author[ibs]{J.~Lee}
\author[knu]{J.~Y.~Lee}
\author[ibs,ust]{M.~H.~Lee}
\author[ibs,ust]{S.~H.~Lee}
\author[snu]{S.~M.~Lee}
\author[cau]{Y.~J.~Lee}
\author[ibs]{D.~S.~Leonard}
\author[knu]{N.~T.~Luan}
\author[usp]{B.~B.~Manzato}
\author[yale]{R.~H.~Maruyama}
\author[shef]{R.~J.~Neal}
\author[yale]{J.~A.~Nikkel}
\author[ibs]{S.~L.~Olsen}
\author[ibs,ust]{B.~J.~Park}
\author[ku]{H.~K.~Park}
\author[kri]{H.~S.~Park}
\author[iqs]{J.~C.~Park}
\author[ibs]{K.~S.~Park}
\author[knu]{S.~D.~Park}
\author[usp]{R.~L.~C.~Pitta}
\author[unm]{H.~Prihtiadi}
\author[ibs]{S.~J.~Ra}
\author[skku,utah]{C.~Rott}
\author[shef]{A.~Scarff}
\author[ibs]{K.~A.~Shin}
\author[iqs]{M.~K.~Son}
\author[shef]{N.~J.~C.~Spooner}
%\author[yale]{W.~G.~Thompson}
\author[knu]{L.~T.~Truc}
\author[ucsd]{L.~Yang}
\author[ibs,skku]{G.~H.~Yu\corref{corr}} \ead{tksxk752@naver.com}
\author[]{(COSINE-100 Collaboration)}

\address[yale]{Department of Physics and Wright Laboratory, Yale University, New Haven, CT 06520, USA}
\address[usp]{Physics Institute, University of S\~{a}o Paulo, 05508-090, S\~{a}o Paulo, Brazil}
\address[ibs]{Center for Underground Physics, Institute for Basic Science (IBS), Daejeon 34126, Republic of Korea}
\address[snu]{Department of Physics and Astronomy, Seoul National University, Seoul 08826, Republic of Korea} 
\address[shef]{Department of Physics and Astronomy, University of Sheffield, Sheffield S3 7RH, United Kingdom}
\address[cau]{Department of Physics, Chung-Ang University, Seoul 06973, Republic of Korea}
\address[cens]{Center for Exotic Nuclear Studies, Institute for Basic Science (IBS), Daejeon 34126, Republic of Korea}
\address[ewu]{Department of Science Education, Ewha Womans University, Seoul 03760, Republic of Korea} 
\address[ust]{IBS School, University of Science and Technology (UST), Daejeon 34113, Republic of Korea}
\address[wipac]{Department of Physics and Wisconsin IceCube Particle Astrophysics Center, University of Wisconsin-Madison, Madison, WI 53706, USA}
\address[knu]{Department of Physics, Kyungpook National University, Daegu 41566, Republic of Korea}
\address[su]{Department of Physics, Sejong University, Seoul 05006, Republic of Korea}
\address[kri]{Korea Research Institute of Standards and Science, Daejeon 34113, Republic of Korea}
\address[ku]{Department of Accelerator Science, Korea University, Sejong 30019, Republic of Korea}
\address[iqs]{Department of Physics and IQS, Chungnam National University, Daejeon 34134, Republic of Korea}
\address[unm]{Department of Physics, Universitas Negeri Malang, Malang 65145, Indonesia}
\address[skku]{Department of Physics, Sungkyunkwan University, Suwon 16419, Republic of Korea}
\address[utah]{Department of Physics and Astronomy, University of Utah, Salt Lake City, UT 84112, USA}
\address[ucsd]{Department of Physics, University of California San Diego, La Jolla, CA 92093, USA}

%% Abstract
\begin{abstract}
COSINE-100 is a dark matter direct detection experiment with 106\,kg NaI(Tl) as the target material. 
$^{210}$Pb and daughter isotopes are a dominant background in the WIMP region of interest and are detected via $\beta$ decay and $\alpha$ decay. 
Analysis of the $\alpha$ channel complements the background model as observed in the \bega channel. 
We present the measurement of the quenching factors and Monte Carlo simulation results and activity quantification of the $\alpha$ decay components of the COSINE-100 NaI(Tl) crystals. 
The data strongly indicate that the $\alpha$ decays probabilistically undergo two possible quenching factors but require further investigation. 
%Furthermore, the $\alpha$ decay energy peaks have highly asymmetric spectral shape. 
%Applying two quenching factors to simulation matches the data well but this phenomena is not understood. 
The fitted results are consistent with independent measurements and improve the overall understanding of the COSINE-100 backgrounds.
Furthermore, the half-life of $^{216}$Po has been measured to be $143.4\,\pm\,1.2$\,ms, which is consistent with and more precise than most current measurements.
\end{abstract}

%% Keywords

\end{frontmatter}

%% enable line numbers
%\linenumbers
%\setlength\linenumbersep{5pt}
%\renewcommand\linenumberfont{\normalfont\tiny\sffamily\color{gray}}

%% Introduction 
\section{Introduction} \label{sec:intro}

The COSINE-100 detector consists of an array of eight ultra-pure NaI(Tl) crystals with a combined mass of 106\,kg. The experiment was operated at the Yangyang Underground Laboratory (Y2L) from September 2016 until March 2023. Its primary goal is the direct detection of dark matter via the annual modulation of WIMP scattering from I or Na nuclei. The crystal array is immersed in 2200 liters of liquid scintillator for background tagging. 

A good understanding and precise measurement of the background sources contributing events in the 1--10\,electron-equivalent\,keV (keV$_{ee}$) region of interest (ROI) is critical to NaI(Tl) direct detection WIMP searches.
Bulk and surface $^{210}$Pb are the dominant background sources in this region and are observed in the \bega channel, as well as in the $\alpha$ channel.
Certain characteristics (gain calibration, energy resolution, light yield, etc.) of the NaI crystals have been measured in situ~\cite{Adhikari:2017esn}.
The quenching factors for $\alpha$ particles are important inputs to the simulation, as well as the analysis of the $\alpha$ background.
%The measured quenching factors are presented in this work.
Analysis of the $\alpha$ channel can improve the accuracy of the background model in the ROI.
Analysis of the $\alpha$ backgrounds using 2.7\,yr data is presented in this work, intended to quantify the activity of the different components by fitting data using Monte Carlo simulations and a direct determination of quenching factors for alpha particles.

%% Setup 
\section{COSINE-100 experimental setup} \label{sec:setup}

\begin{figure}[ht] \centering
\includegraphics[width=0.9\linewidth]{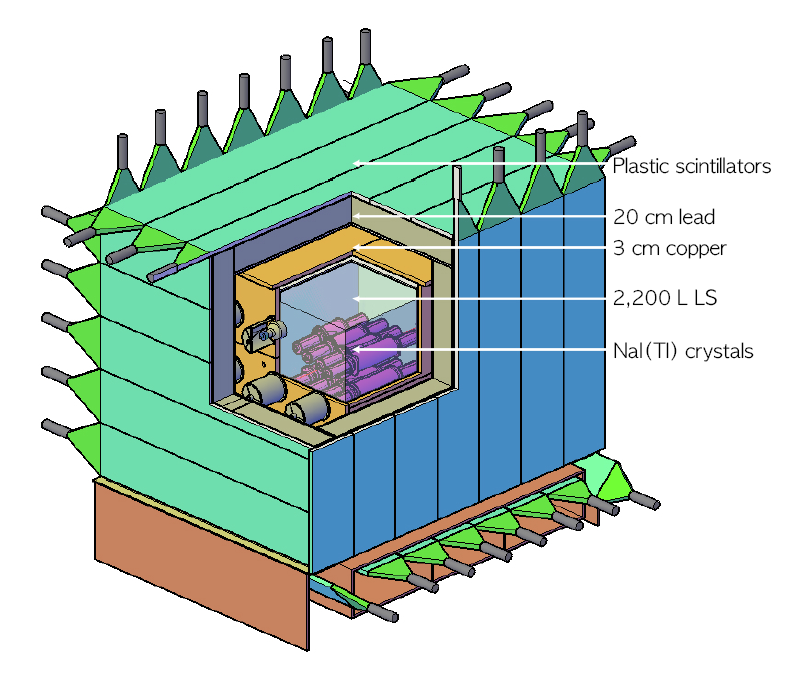}
\caption{COSINE-100 detector geometry implemented in Geant4 (Sec.~\ref{sec:modeling}).} 
\label{fig:G4det}
\end{figure}

The experimental setup and data acquisition are described in detail in Ref.~\cite{Adhikari:2017esn, COSINE-100:2018rxe}. The detector geometry used for simulations is shown in Fig.~\ref{fig:G4det}. Eight NaI(Tl) crystals (denoted C1-C8), arranged in two layers, are located in the middle of a four-layer shielding structure. From outside inward, the four shielding layers are plastic scintillator panels, a lead-brick castle, a copper box, and a scintillating liquid. The eight NaI(Tl) crystal assemblies and their support table are immersed in the scintillating liquid that serves both as an active veto and a passive shield \cite{Adhikari:2020asl, COSINE-100:2017dsl}. 

The eight NaI(Tl) crystals were grown out of batches of powder provided by Alpha Spectra Inc. with successive improvements. 
The first crystals labeled AS-B and AS-C were produced with a reduction of $^{40}$K by an order of magnitude compared to crystals grown with AS-A powder.
%The first samples, which produced an order of magnitude reduction in $^{40}$K, were AS-B and AS-C. 
Then, crystals labeled WIMPScint-II (AS-WSII) reduced the contamination of $^{210}$Pb.
%This was followed by WIMPScint-II (AS-WSII) which reduced the $^{210}$Pb contamination, and WIMPScint-III (AS-WSIII) which resulted in another factor of two reduction of $^{40}$K. 
Finally, the crystals labeled WIMPScint-III (AS-WSIII) were produced with an additional reduction of $^{40}$K by a factor of two. 
The crystal properties are summarized in Table~\ref{tab:crystal-overview}, and a detailed description is given in Ref.~\cite{Adhikari:2017esn}. 

The final crystals are cylindrically shaped and hermetically encased in OFE copper tubes with wall thickness of 1.5\,mm and quartz windows (12.0\,mm thick) at each end. Each crystal's lateral surface is wrapped in roughly 10 layers of 250\,$\mu$m thick PTFE reflective sheets. The quartz windows are optically coupled to each end of the crystal via 1.5\,mm thick optical pads. These, in turn, are optically coupled to 3-inch Hamamatsu R12669SEL photomultiplier tubes (PMTs) with a thin layer of high-viscosity optical gel. The PMTs are sealed from the liquid scintillator by a housing made of copper and PTFE.

\begin{table*}[ht] \centering
\begin{tabular}{cccccc} 
NaI(Tl)  & Mass  & Size (inches)     & Powder   & Total $\alpha$ Rate   & $^{40}$K  \\
Crystal  & (kg)  & (diam. $\times$ length)  & Type     & (mBq/kg)              & (ppb)     \\
\hline 
C1    &  8.3    &  5.0 x 7.0    & AS-B      & 3.20 $\pm$ 0.08   & 34.7 $\pm$ 4.7  \\
C2    &  9.2    &  4.2 x 11.0   & AS-C      & 2.06 $\pm$ 0.06   & 60.6 $\pm$ 4.7  \\
C3    &  9.2    &  4.2 x 11.0   & AS-WSII   & 0.76 $\pm$ 0.02   & 34.3 $\pm$ 3.1  \\
C4    &  18.0   &  5.0 x 15.3   & AS-WSII   & 0.74 $\pm$ 0.02   & 33.3 $\pm$ 3.5  \\
C5    &  18.3   &  5.0 x 15.5   & AS-C      & 2.06 $\pm$ 0.05   & 82.3 $\pm$ 5.5  \\
C6    &  12.5   &  4.8 x 11.8   & AS-WSIII  & 1.52 $\pm$ 0.04   & 16.8 $\pm$ 2.5  \\
C7    &  12.5   &  4.8 x 11.8   & AS-WSIII  & 1.54 $\pm$ 0.04   & 18.7 $\pm$ 2.8  \\
C8    &  18.3   &  5.0 x 15.5   & AS-C      & 2.05 $\pm$ 0.05   & 54.3 $\pm$ 3.8  \\
\end{tabular} 
\caption{COSINE-100 crystal properties. The total $\alpha$ rates were independently measured as part of the crystal commissioning effort. Details of the measurements are described in Ref.~\cite{Adhikari:2017esn}.}
\label{tab:crystal-overview}
\end{table*}

It should be noted that Crystal-5 and Crystal-8 are excluded from this work due to significant degradation of the light collection. 
%These two crystals exhibit significantly decreased light collection. Degradation of the optical coupling between the crystal and PMTs is suspected. The poor light collection leads to an imprecise data calibration and MC energy resolution correction. 

The scintillation signals are digitized and integrated to determine the charge in arbitrary units. Special calibration runs and prominent peaks from in situ data are used to define the calibration function~\cite{COSINE-100:2018rxe} that converts the digitized charge into units of electron-volts (eV). The calibration function is determined from the \bega scintillation signals and is referred to as the electron-equivalent energy (eV$_{ee}$).

%% Quenching 
\section{Alpha quenching measurements} \label{sec:quenching}

The $\alpha$ events are selected from data using the meantime variable as discussed in Sec.~\ref{sec:alpha-anal}. The observed energy of the $\alpha$ events is used to build the energy spectra. The energy spectra are in units of electron-equivalent energy and do not represent the true $\alpha$ energy. Only a fraction of the $\alpha$ decay energy is converted to light. The ratio of the observed energy to the decay energy is referred to as the $\alpha$ quenching factor (QF). To determine the crystal quenching factors, known $\alpha$ peaks are identified in data as described in Sec.~\ref{sec:aa-events}. The quenching factor is also $\alpha$ energy dependent so ideally multiple known $\alpha$ peaks are required to determine the energy dependent quenching factor function as summarized in Sec.~\ref{sec:q-factors}.

%\subsection{Alpha analysis} \label{sec:alpha-anal}
\subsection{Alpha event selection and energy spectra} \label{sec:alpha-anal}

\begin{figure}[ht] \centering
\includegraphics[width=0.9\linewidth]{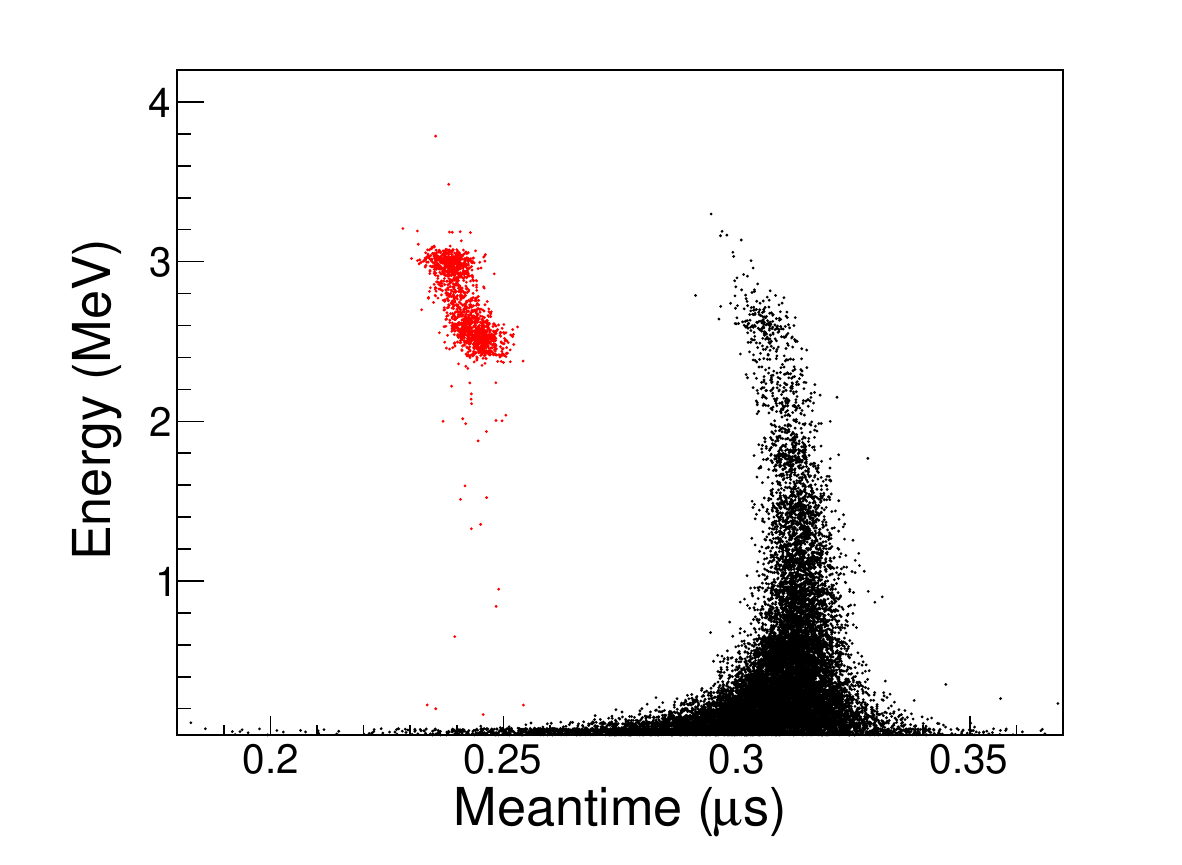}
\caption{Typical crystal scatter plot of the meantime versus the energy for COSINE-100 background data \capt. The $\alpha$ events (red dots) and the \bega events (black dots) are clearly identified by the meantime as defined by Eq.~\ref{eq:meantime}.}
\label{fig:AlphaScatter}
\end{figure}

%Alpha signals in the NaI(Tl) crystals can be separated from \bega events by using fast scintillation decay time of incident $\alpha$ particles. 
The scintillation signal from $\alpha$ particles in NaI(Tl) differs from the one generated by \bega particles.
The decay time of the scintillation pulse by $\alpha$ is faster than the one from \bega particles. 
The scintillation decay time of the incident particles can be identified from the charge-weighted duration time, called the meantime. The meantime $\langle t \rangle$ is defined as
\begin{eqnarray}
\langle t \rangle = \frac{\Sigma_{i} A_{i} t_{i}}{\Sigma_{i} A_{i}} ,
\label{eq:meantime}
\end{eqnarray}
where $A_{i}$ and $t_{i}$ are the charge and time of the $\textit{i}$-th digitized bin of a signal waveform, respectively. 
%The meantime is estimated within 1.5\,$\mu$s from the pulse start time. 
The meantime is estimated in COSINE-100 data within 1.5\,$\mu$s from the pulse start time. 
As a result, the populations of \bega events and $\alpha$ events are clearly separated due to the faster decay times of the $\alpha$-induced events as shown in Fig.~\ref{fig:AlphaScatter}.

\begin{figure}[ht] \centering
\includegraphics[width=0.9\linewidth]{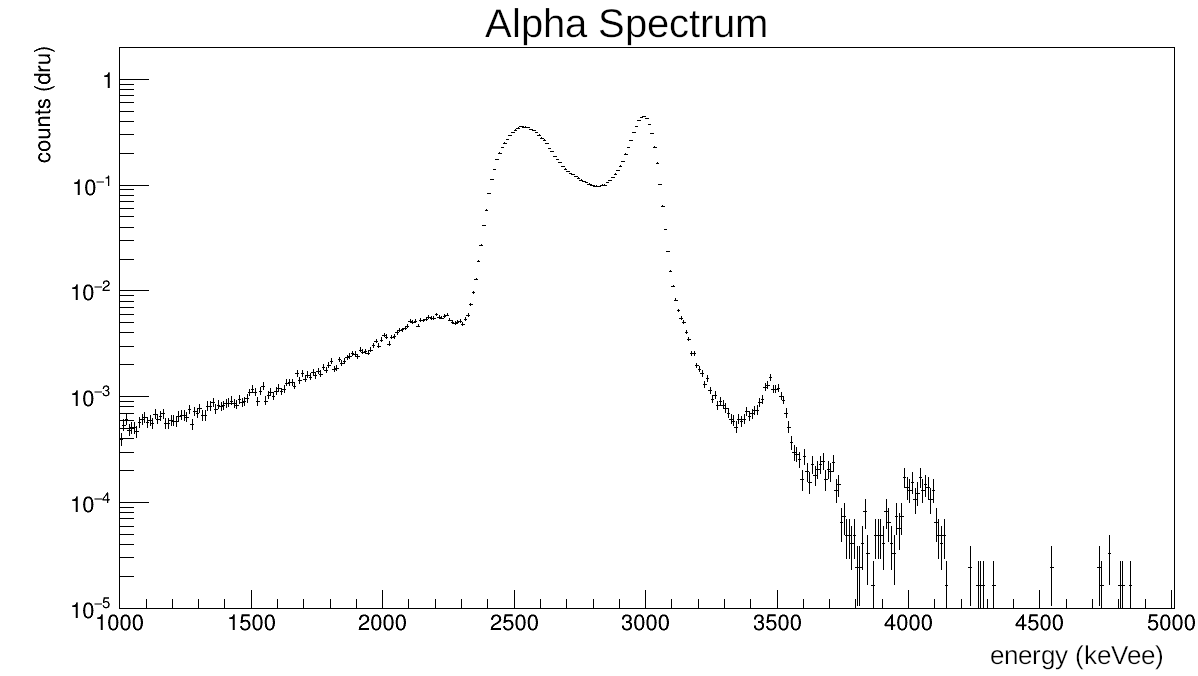}
\caption{Characteristic electron-equivalent energy spectrum of $\alpha$ events \capt. $^{210}$Po makes up the two prominent peaks suggesting two quenching factors.}
\label{fig:just-data}
\end{figure}

%\subsection{Electron-equivalent alpha energy spectra} \label{sec:alpha-spectra}
Using the meantime selection criteria, the $\alpha$ events are selected from data. We obtain the $\alpha$ energy spectra in units of electron-equivalent energy (keV$_{ee}$) as shown in Fig.~\ref{fig:just-data}. 
%Only a fraction of the $\alpha$ decay energy is converted to light and is referred to as the $\alpha$ quenching factor (QF). 
The half-lives of the two prominent peaks are consistent with $^{210}$Po. 
The ANAIS experiment observed two prominent peaks in their alpha spectra~\cite{Amare:2016rbf} and cautiously proposes surface $^{210}$Po as an explanation. 
Discussed in detail in Sec.~\ref{sec:modeling}, MC simulation of surface $^{210}$Po was insufficient to explain the two peak structure observed in our data.
A recent preliminary study by the COSINUS experiment~\cite{Bharadwaj:2023} may provide stronger evidence of QF dependence on Tl doping concentration once the study matures. 
In principle, the NaI(Tl) crystal growing processes can cause spatial dependence of the Tl doping concentration within the crystal. The spatial dependence of the QF could be a continuum, distinctly different, or a combination of both. 
In this work, we consider two different quenching factors for $\alpha$ events. The high QF and low QF are referred to as $Q_{1}$ and $Q_{2}$ respectively.

%\begin{figure*}[ht] 
%\centering
%\includegraphics[width=0.9\textwidth]{figs/just-alpha-data.pdf}
%\caption{The $\alpha$ electron-equivalent energy data spectra of the COSINE-100 crystals. $^{210}$Po makes up the prominent peaks following two quenching factors with asymmetric shape.}
%\label{fig:AlphaEnergy}
%\end{figure*}

%\begin{figure*}[ht]
%\centering
%\subfloat[Energy spectrum of C2]{
%  \includegraphics[width=0.19\textwidth]{./figs/C2_Alpha.pdf}\label{fig:C2Energy}}
%\subfloat[Energy spectrum of C3]{
%  \includegraphics[width=0.19\textwidth]{./figs/C3_Alpha.pdf}\label{fig:C3Energy}}
%\subfloat[Energy spectrum of C4]{
%  \includegraphics[width=0.19\textwidth]{./figs/C4_Alpha.pdf}\label{fig:C4Energy}}
%\subfloat[Energy spectrum of C6]{
%  \includegraphics[width=0.19\textwidth]{./figs/C6_Alpha.pdf}\label{fig:C6Energy}}
%\subfloat[Energy spectrum of C7]{
%  \includegraphics[width=0.19\textwidth]{./figs/C7_Alpha.pdf}\label{fig:C7Energy}}
%\caption{The total $\alpha$ energy spectra of COSINE-100 crystals. $^{210}$Po makes the highest peaks following unexpected shape in contrast to $\gamma$/$\beta$ events.}
%\label{fig:AlphaEnergy}
%\end{figure*}

\subsection{Alpha-alpha time-correlated events} \label{sec:aa-events}
The high energy region ($>$\,3.3\,\mevee) of the $\alpha$ distribution, as shown in Fig.~\ref{fig:just-data}, is from the $^{228}$Th-group $\alpha$ decays and more specifically from $^{220}$Rn and $^{216}$Po. The $^{228}$Th-group activity can be measured by selecting the time-delayed \alal events from $^{220}$Rn and $^{216}$Po decays. The $\alpha$ decay of $^{216}$Po has a short half-life of 0.145\,s following its production via the $\alpha$ decay of $^{220}$Rn. 
%In spite of this short half-life, the activity of $^{210}$Po is about 10$^{4}$ times higher than the $^{228}$Th-group and the accidental coincidence rate is high. Therefore, we apply an energy constraint to mitigate random coincident events from $^{210}$Po.

\begin{figure}[ht] 
\centering
\includegraphics[width=0.9\linewidth]{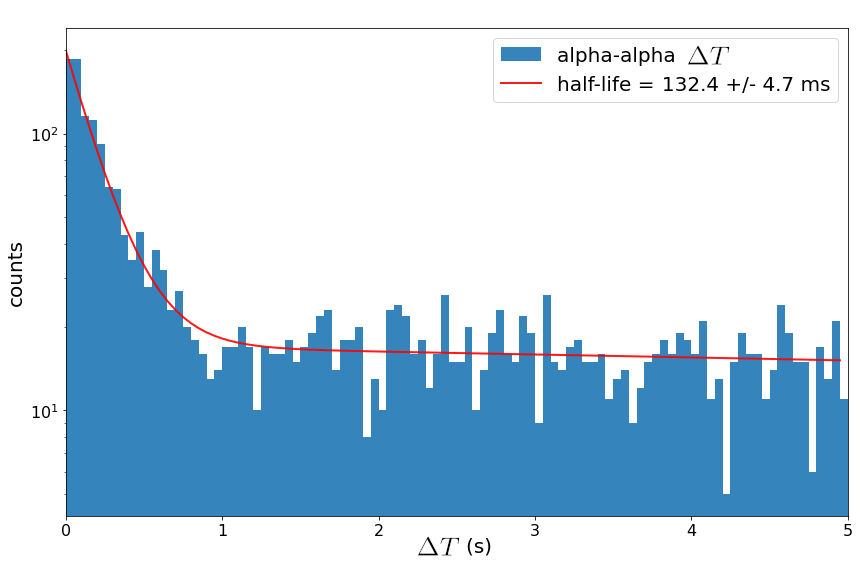}
\caption{A typical \dt distribution of total $\alpha$ events and exponential fitted function \capt.}
\label{fig:alpha-coinc}
\end{figure}

The $^{228}$Th-group activity can be calculated from a single exponential fit to the \alal time-delay (\dt) distribution. 
%An energy cut around the $^{216}$Po energy peak position is applied to reduce random coincident events from $^{210}$Po.
%while maintaining high $^{220}$Rn and $^{216}$Po selection efficiency. 
Events with \dt greater than 5 seconds are rejected. 
%The relatively long \dt provides more events for the $^{210}$Po background estimation.
A typical \dt distribution and fitted function is shown in Fig.~\ref{fig:alpha-coinc} and the fitted half-lives of all crystals are summarized in Table~\ref{tab:my-half-lifes}. 
%The \alal coincidence measurements are compared to fitted $\alpha$ spectra results in Table~\ref{tab:th228-group-activities}.
%The $^{228}$Th activities measured in the \alal coincidence study are listed in Table~\ref{tab:th228-group-activities}.

%\begin{figure*}[ht] \centering
%\subfloat{\includegraphics[width=0.3\textwidth]{figs/C1_TimeGap.pdf} \label{fig:C1TimeGap}}
%\subfloat{\includegraphics[width=0.3\textwidth]{figs/C2_TimeGap.pdf} \label{fig:C2TimeGap}}
%\subfloat{\includegraphics[width=0.3\textwidth]{figs/C3_TimeGap.pdf} \label{fig:C3TimeGap}}
%\\
%\subfloat{\includegraphics[width=0.3\textwidth]{figs/C4_TimeGap.pdf} \label{fig:C4TimeGap}}
%\subfloat{\includegraphics[width=0.3\textwidth]{figs/C6_TimeGap.pdf} \label{fig:C6TimeGap}}
%\subfloat{\includegraphics[width=0.3\textwidth]{figs/C7_TimeGap.pdf} \label{fig:C7TimeGap}}
%\caption{The \dt distributions of total $\alpha$ events and exponential fitted functions.}
%\label{fig:alpha-coinc}
%\end{figure*}

\begin{table}[ht] \centering
\begin{tabular}{ c | c }
      &  Half-lives of $^{216}$Po ($ms$) \\
\hline 
C1    &  150.8 $\pm$ 4.7  \\
C2    &  141.3 $\pm$ 1.5  \\
C3    &  133.8 $\pm$ 5.1  \\
C4    &  152.8 $\pm$ 2.6  \\
C6    &  132.4 $\pm$ 4.7  \\
C7    &  148.1 $\pm$ 6.8  \\
\end{tabular} 
\caption{Measured $^{216}$Po half-lives (in $ms$) from exponential fit to \alal time distribution.}
\label{tab:my-half-lifes}
\end{table}

It should be noted that our measured average $^{216}$Po half-life of $143.4\,\pm\,1.2\,ms$ is very competitive with the leading measurements~\cite{Danevich:2003, Naddard:2017, Azzolini:2021, Diamond:1963, Baccolo:2021}. This measurement is more precise than previous measurements with the exception of~\cite{Naddard:2017}. Previous $^{216}$Po half-life measurements are listed in Table~\ref{tab:po216-hls} for comparison.

\begin{table}[ht] \centering
\begin{tabular}{ c | c }
    &  Half-lives of $^{216}$Po ($ms$) \\
    \hline 
    this work            &  143.4 $\pm$  1.2  \\
    \cite{Danevich:2003} &  144   $\pm$  8    \\
    \cite{Naddard:2017}  &  144.0 $\pm$  0.6  \\
    \cite{Azzolini:2021} &  143.3 $\pm$  2.8  \\
    \cite{Diamond:1963}  &  145   $\pm$  2    \\
    \cite{Baccolo:2021}  &  145.3 $\pm$ 18.9  \\
    \hline 
    global average       &  143.9 $\pm$  0.5  \\
\end{tabular} 
\caption{Measured $^{216}$Po half-lives (in $ms$) from other sources.}
\label{tab:po216-hls}
\end{table}

\subsection{Quenching factor calculation} \label{sec:q-factors}
\begin{figure}[ht] \centering
\subfloat{\includegraphics[width=0.8\linewidth]{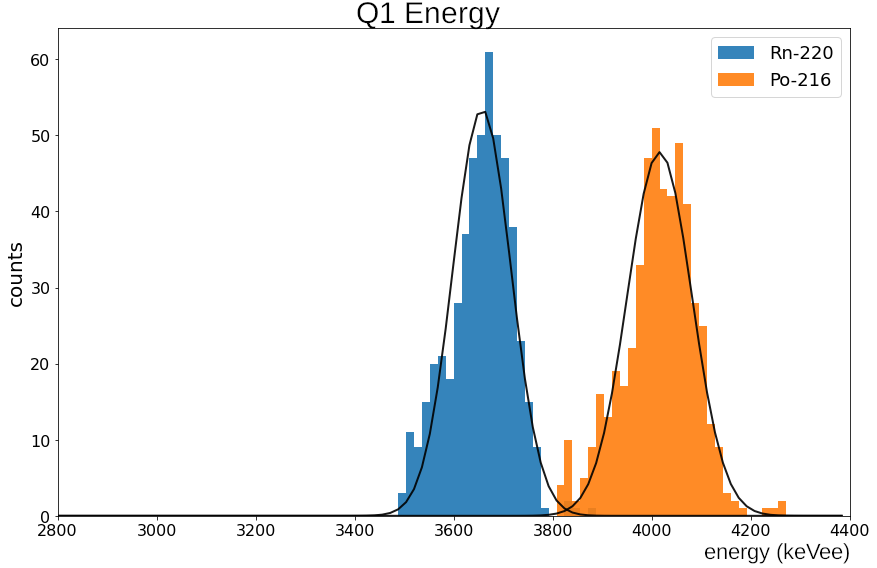} \label{fig:Po-Rn-Q1}}
\\
\subfloat{\includegraphics[width=0.8\linewidth]{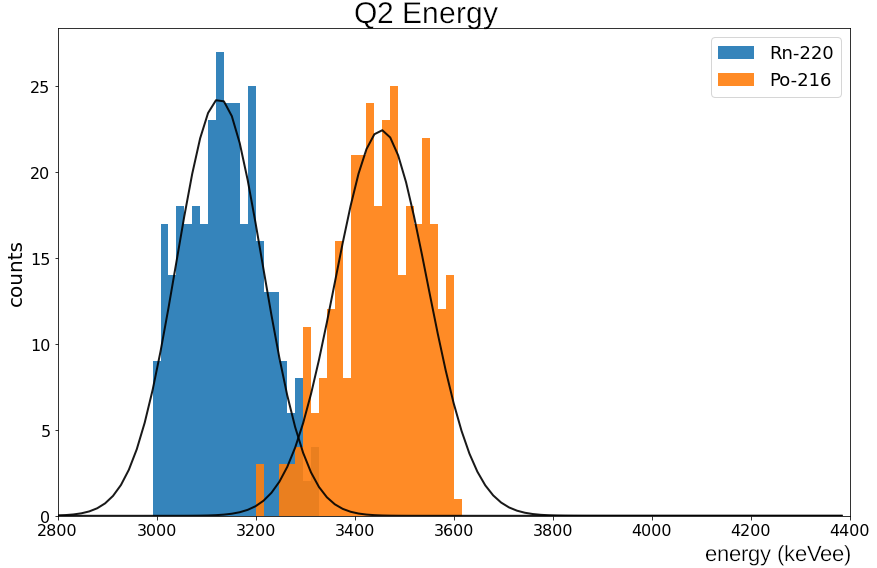} \label{fig:Po-Rn-Q2}}
\caption{Characteristic \alal electron-equivalent energy distributions of the selected $^{216}$Po and $^{220}$Rn events for $Q_{1}$ and $Q_{2}$ \capt.}
\label{fig:alpha-alpha-energy}
\end{figure}

%The peak positions of $^{220}$Rn and $^{216}$Po are determined from the \alal coincidence energy spectra. Strict energy and \dt constraints are applied to mitigate random coincidence events from $^{210}$Po.
The $^{220}$Rn and $^{216}$Po events that have been selected are shown in Fig.~\ref{fig:alpha-alpha-energy}. 
These events were estimated using the time-delayed method outlined in Section~\ref{sec:aa-events}. 
They are used to determine the energy-dependent quenching functions.
The observed $\alpha$ energy and standard deviation of $^{220}$Rn and $^{216}$Po events are initially estimated by Gaussian fits to the identified spectral features in data. 
The fitted parameters are used to define the energy window applied to the \alal event selection. 
The energy window is set by the $\alpha$ energy mean plus or minus three standard deviations. 
A time constraint is also applied to mitigate random coincident events from $^{210}$Po. 
From the half-life of $^{216}$Po, 0.145\,s, we apply a 1\,s \dt constraint. 
%When a prompt $\alpha$ event satisfies the $^{220}$Rn energy criteria, if the following $\alpha$ event satisfies the $^{216}$Po energy criteria and the \dt constraint is satisfied, the two events are tagged as an \alal time-correlated event. 
%The time distribution of the \alal events is used to verify the half-life and proper identification of $^{216}$Po. 
%The resulting electron-equivalent mean energies of the selected $^{220}$Rn and $^{216}$Po events, as seen in Fig.~\ref{fig:alpha-alpha-energy}, are used to determine the energy dependent quenching functions.

%\begin{figure}[ht] \centering
%\includegraphics[width=0.9\linewidth]{figs/c4_q1_energy.png}
%\caption{Characteristic \alal electron-equivalent energy distributions of the selected $^{220}$Rn and $^{216}$Po events. } 
%\label{fig:alpha-alpha-energy}
%\end{figure}

\begin{figure}[ht] \centering
\includegraphics[width=0.9\linewidth]{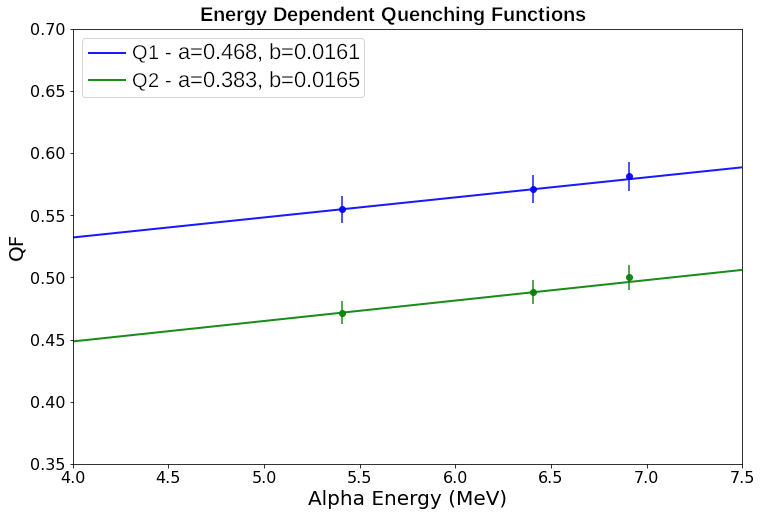}
\caption{Characteristic data points and energy-dependent quenching factor functions for $Q_{1}$ and $Q_{2}$ \capt.} 
\label{fig:q1q2-fits}
\end{figure}

The energy-dependent $\alpha$ quenching is approximated by
\begin{eqnarray}
 Q_{\alpha}(E_{\alpha}) = a + b \cdot E_{\alpha}[\text{MeV}]\,,
\label{eq:quenching}
\end{eqnarray}
and each data point for fitting is obtained from the ratio between the Q-value and the electron-equivalent energy of $\alpha$ from $^{210}$Po, $^{220}$Rn, and $^{216}$Po. Characteristic data points and energy-dependent quenching factor functions for $Q_{1}$ and $Q_{2}$ are shown in Fig.~\ref{fig:q1q2-fits}. The fitted $a$ and $b$ parameters for each crystal are summarized in Tables~\ref{tab:qf1-pars} and~\ref{tab:qf2-pars}.

%\begin{table}[ht] \centering
%\begin{tabular}{ c | c c | c c }
%      & \multicolumn{2}{c}{$Q_{1}$}  & \multicolumn{2}{c}{$Q_{2}$} \\
%      &  $a$    &  $b$     &  $a$    &  $b$  \\
%\hline 
%C1    &  0.437  &  0.0218  &  -  &  -  \\
%C2    &  0.465  &  0.0210  &  0.415  &  0.0216  \\
%C3    &  0.419  &  0.0242  &  0.394  &  0.0238  \\
%C4    &  0.458  &  0.0214  &  0.425  &  0.0223  \\
%C6    &  0.468  &  0.0161  &  0.383  &  0.0165  \\
%C7    &  0.493  &  0.0114  &  0.399  &  0.0135  \\
%\end{tabular} 
%\caption{Energy dependent $\alpha$ quenching factor parameters as defined by Eq.~\ref{eq:quenching}.}
%\label{tab:qf-pars}
%\end{table}

\begin{table}[ht] \centering
\begin{tabular}{ c | c c }
      & \multicolumn{2}{c}{$Q_{1}$} \\
      &  $a$    &  $b$     \\
\hline 
C1    &  0.437 $\pm$ 0.008  &  0.0218 $\pm$ 0.0014 \\
C2    &  0.465 $\pm$ 0.008  &  0.0210 $\pm$ 0.0013 \\
C3    &  0.419 $\pm$ 0.008  &  0.0242 $\pm$ 0.0014 \\
C4    &  0.458 $\pm$ 0.008  &  0.0214 $\pm$ 0.0013 \\
C6    &  0.468 $\pm$ 0.008  &  0.0161 $\pm$ 0.0013 \\
C7    &  0.493 $\pm$ 0.007  &  0.0114 $\pm$ 0.0012 \\
\end{tabular} 
\caption{Energy-dependent $Q_{1}$ parameters as defined by Eq.~\ref{eq:quenching} obtained for each crystal.}
\label{tab:qf1-pars}
\end{table}

\begin{table}[ht] \centering
\begin{tabular}{ c | c c }
      & \multicolumn{2}{c}{$Q_{2}$} \\
      &  $a$    &  $b$  \\
\hline 
C1    &  0.274 $\pm$ 0.011 &  0.0425 $\pm$ 0.0018 \\
C2    &  0.415 $\pm$ 0.009 &  0.0216 $\pm$ 0.0014 \\
C3    &  0.394 $\pm$ 0.009 &  0.0238 $\pm$ 0.0015 \\
C4    &  0.425 $\pm$ 0.008 &  0.0223 $\pm$ 0.0014 \\
C6    &  0.383 $\pm$ 0.009 &  0.0165 $\pm$ 0.0015 \\
C7    &  0.399 $\pm$ 0.009 &  0.0135 $\pm$ 0.0015 \\
\end{tabular} 
\caption{Energy-dependent $Q_{2}$ parameters as defined by Eq.~\ref{eq:quenching} obtained for each crystal.}
\label{tab:qf2-pars}
\end{table}

%% Modeling 
\section{Alpha decay Monte Carlo simulation and activity determination} \label{sec:modeling}

Radioactive background components are simulated using Geant4 (v10.4.2)~\cite{GEANT4:2002zbu, Allison:2006ve, Allison:2016lfl}. The physics list classes of G4EmLivermorePhysics for low-energy electromagnetic processes and G4RadioactiveDecay for radioactive decay processes are used. Details of the simulation are described in Refs.~\cite{Adhikari:2017gbj, COSINE-100:2018tfl}.

\begin{figure}[ht] \centering
\includegraphics[width=0.9\linewidth]{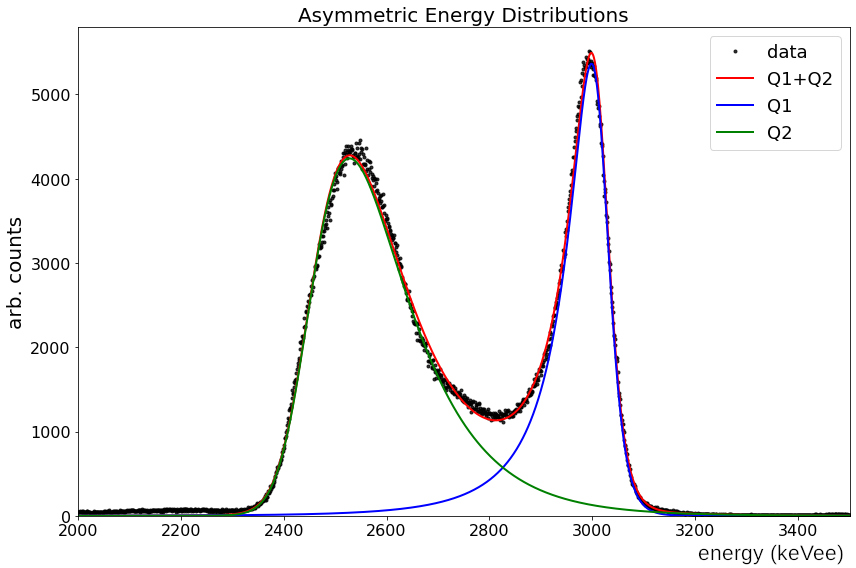}
\caption{Internal $^{210}$Po data approximated by simulation of two quenching factors with normal-inverse Gaussian energy distributions \capt.}
\label{fig:asym-shape}
\end{figure}

The $\alpha$ events are challenging to reproduce with simulation. Two quenching factors are required to describe the two dominant $\alpha$ peaks from $^{210}$Po as well as the higher energy features from $^{220}$Rn and $^{216}$Po. Furthermore, the $^{210}$Po spectral shapes are highly asymmetric, and an asymmetric probability density function (PDF) is required for the simulation energy smearing. One hypothesis is that the asymmetric shape is caused by spatial QF transitioning regions in the NaI(Tl) crystal. Lacking a physics explanation for the asymmetric energy distribution, a PDF was selected that closely approximated the spectral shape observed in data. In this work, the spectral shapes were approximated by the normal-inverse Gaussian~\cite{Barndorff:1977} as demonstrated in Fig.~\ref{fig:asym-shape}. 

The two prominent peaks with asymmetric spectral shape are also observed by the ANAIS experiment~\cite{Amare:2016rbf}, which uses NaI(Tl) crystals produced by Alpha Spectra Inc., the same vendor that produced the crystals for COSINE-100. 
The ANAIS experiment cautiously proposes the lower energy peak could be due to surface $^{210}$Po. We did investigate NaI(Tl) surface $^{210}$Po at depths ranging from $10\,\mu m$ to $10\,nm$ using Geant4 MC simulation but could not demonstrate the large energy difference ($300-500\,keV_{ee}$) observed in data. We also investigated the effects of a scintillation dead layer~\cite{Yang:2014} but could not demonstrate large energy differences. At most, energy differences while considering surface $^{210}$Po and/or a dead layer were $50-100\,keV_{ee}$. 
We hypothesized in Sec.~\ref{sec:alpha-anal} the double peak structure could be explained by the spatial dependence of Tl doping concentration causing two quenching factors, but further investigation is needed.

%The two quenching factors and asymmetric spectral shapes are also observed by the ANAIS experiment~\cite{Amare:2016rbf}, which uses NaI(Tl) crystals produced by Alpha Spectra Inc., the same vendor that produced the crystals for COSINE-100. 
%It is believed that the thallium doping and/or crystal growing procedure causes two different quenching factors. 

%We explored alternative explanations for the double-peak structure in $\alpha$ data. Using MC, we investigated NaI surface $^{210}$Pb at various depths considering uniform concentrations and exponentially weighted concentrations. We investigated the effects of a scintillation dead layer~\cite{Yang:2014} extending to various depths in combination with NaI surface $^{210}$Pb. None of these scenarios sufficiently describe the observed two peak structure in data. Internal and surface $^{210}$Pb have significantly different spectral shapes in the \bega channel and previous measurements of internal $^{210}$Pb~\cite{COSINE-100:2018tfl} are consistent with the combined activity of the two $^{210}$Po peaks in $\alpha$ data. 

The energy dependent $\alpha$ quenching functions determined in Sec.~\ref{sec:q-factors} are separately applied to the simulated $\alpha$ decay events to create $Q_{1}$ and $Q_{2}$ quenched energy distributions. The quenched energies are then smeared asymmetrically. The resulting simulated $\alpha$ spectra are in units of electron-equivalent energy and make up the templates that are fit to data. The fractional activities of the $Q_{1}$ and $Q_{2}$ templates are left as free parameters of the fit. The fitted fractional activities for $^{210}Po$ and $^{228}Th$ are listed in Table~\ref{tab:fractional-activities}. 
For other fitted isotopes, poor statistics does not allow to distinguish $Q_{1}$ and $Q_{2}$, so the fractional activities are assumed to be equal. 
%The other fitted isotopes are assumed to have equal $Q_{1}$ and $Q_{2}$ because there are no distinguishing features in the energy spectra. 
%It is interesting to note different isotopes display different fractions. 

\begin{table}[ht] \centering
	\begin{tabular}{ c | c c | c c }
		& \multicolumn{2}{c}{$^{210}Po$} & \multicolumn{2}{c}{$^{228}Th$} \\
		& \multicolumn{2}{c}{($mBq/kg$)} & \multicolumn{2}{c}{($\mu Bq/kg$)} \\
		&  $Q_{1}$    &  $Q_{2}$  &  $Q_{1}$    &  $Q_{2}$  \\
		\hline 
		C1    & 1.55 & 1.38 & 1.84 & 5.63 \\
		C2    & 0.48 & 1.38 & 2.10 & 2.46 \\
		C3    & 0.22 & 0.41 & 0.75 & 1.16 \\
		C4    & 0.27 & 0.41 & 1.41 & 1.52 \\
		C6    & 0.82 & 0.81 & 0.34 & 1.76 \\
		C7    & 1.01 & 0.62 & 0.41 & 1.73 \\
	\end{tabular} 
	\caption{Fitted fractional activities of $^{210}Po$ and $^{228}Th$ for each crystal. 
 For other fitted isotopes, poor statistics does not allow to distinguish $Q_{1}$ and $Q_{2}$, so the fractional activities are assumed to be equal.
 %The other fitted isotopes are assumed to have equal $Q_{1}$ and $Q_{2}$ because there are no distinguishing features in the energy spectra. 
 }
	\label{tab:fractional-activities}
\end{table}

\subsection{Data fitting} \label{sec:fitting}
\begin{figure*}[ht] \centering
	\includegraphics[width=0.9\textwidth]{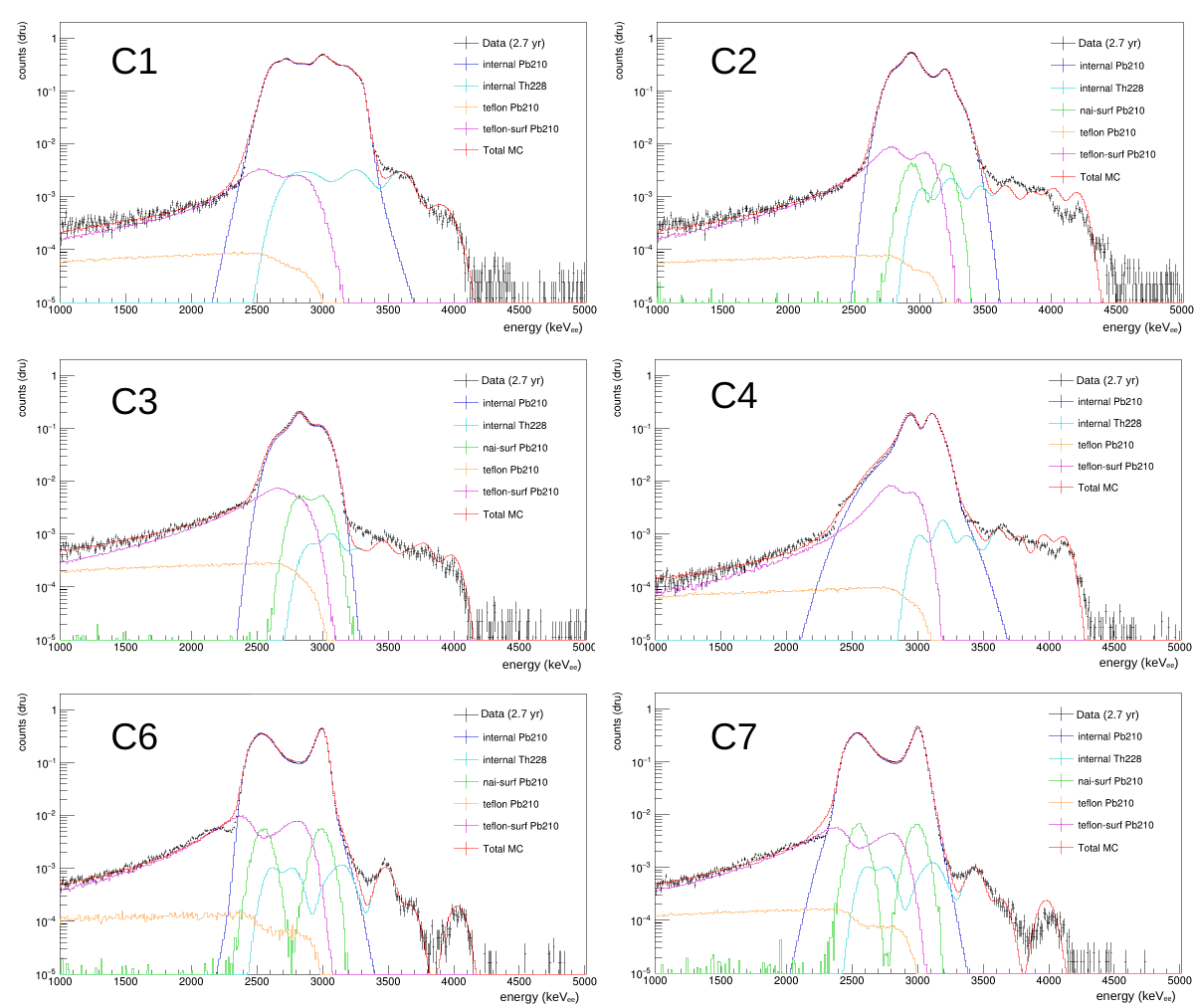}
	\caption{Results of fitting internal $^{210}$Po, internal $^{228}$Th-group, and PTFE surface $^{210}$Po $\alpha$ components to data for each COSINE-100 crystal.}
	\label{fig:fitted-results}
\end{figure*}

Data are fitted by analytically maximizing the binned likelihood of two or more MC background templates as described in Ref.~\cite{Barlow:1993dm}. For each $\alpha$ component, MC templates are created for two quenching functions and allowed to be independently scaled by the fitting algorithm. The resulting fits to the COSINE-100 crystals are shown in Fig.~\ref{fig:fitted-results}. 
%The fitted activities of the dominant $\alpha$ components are listed in Table~\ref{tab:fitted-activities}. 

The activities of less significant components of PTFE bulk $^{210}$Po and NaI surface $^{210}$Po are determined from analysis of the \bega channel as presented in Ref.~\cite{COSINE-100:2018tfl}. These background sources are included in the $\alpha$ background model but their activities are fixed for this work.

%Crystal-5 and Crystal-8 are excluded from this work. These two crystals exhibit significant degradation of light yield and/or light collection. Their energy resolutions are insufficient for identifying the spectral features required for accurately calibrating the data and scaling the energy resolution of the MC. 

\subsection{Spectral components} \label{sec:features}
The $\alpha$ spectra are dominated by three distinct components. They include internal $^{210}$Po (2.2--3.2\,MeV$_{ee}$), internal $^{228}$Th-group (3.2--4.2\,MeV$_{ee}$), and PTFE surface $^{210}$Po (1.0--2.2\,MeV$_{ee}$). PTFE bulk $^{210}$Po and NaI surface $^{210}$Po are included in the background model but do not contribute significantly to the $\alpha$ spectrum. The activities of these components are measured in the \bega channel and a detailed analysis is presented in Ref.~\cite{COSINE-100:2018tfl}. The fitted activities of NaI surface $^{210}$Po in the \bega channel in crystals C1 and C4 were consistent with the null hypothesis, so this component is not included in the $\alpha$ spectra of C1 and C4.

\subsubsection{Internal \texorpdfstring{$^{210}Po$}{210Po}} \label{sec:internal-po210}
The $\alpha$ spectra are dominated by internal $^{210}$Po at 2.2--3.2\,MeV$_{ee}$. %Geant4 simulated $^{210}$Po $\alpha$ events are used to verify the peak positions and quenching factors in data. The energy distortion of the peaks in data is not understood. NaI surface $^{210}$Po and dead-layer effects were investigated but did not improve the agreement between simulation and data. Each crystal displays different distortion characteristics. We speculate the distortion could be caused by varying Tl concentrations, channeling effects, or something else. 
It is not understood why the $^{210}$Po $\alpha$ energy distribution is highly asymmetric. 
For the purpose of this $\alpha$ modeling, the $^{210}$Po spectral shape is parameterized by normal-inverse Gaussian distributions as this approach best approximates the shape of the $^{210}$Po spectrum observed in data. 
Internal $^{210}$Po is the dominant $\alpha$ background and Table~\ref{tab:internal-po210-activities} compares the fitted activities to measurements of the total $\alpha$ rates~\cite{Adhikari:2017esn}.

\begin{table}[ht] \centering
	\begin{tabular}{ c | c | c } 
		& Fitted MC           & Total            \\
		& $^{210}$Po          & $\alpha$ Rate    \\
		& ($mBq/kg$)          & ($mBq/kg$)       \\
		\hline 
		C1    & 2.93 $\pm$ 0.17   & 3.20 $\pm$ 0.08  \\
		C2    & 1.86 $\pm$ 0.17   & 2.06 $\pm$ 0.06  \\
		C3    & 0.65 $\pm$ 0.10   & 0.76 $\pm$ 0.02  \\
		C4    & 0.68 $\pm$ 0.11   & 0.74 $\pm$ 0.02  \\
		C6    & 1.62 $\pm$ 0.14   & 1.52 $\pm$ 0.04  \\
		C7    & 1.63 $\pm$ 0.14   & 1.54 $\pm$ 0.04  \\
	\end{tabular}
	\caption{Comparison of MC fitted activity of internal $^{210}$Po to the total $\alpha$ rate measured in Ref.~\cite{Adhikari:2017esn}.}
	\label{tab:internal-po210-activities}
\end{table}

\subsubsection{Internal \texorpdfstring{$^{228}$Th}{228Th}-group} \label{sec:internal-th232}
The features visible at higher energies ($>$\,3.2\,MeV$_{ee}$) are dominated by internal $^{232}$Th and particularly the $^{228}$Th-group consisting of $^{228}$Th, $^{224}$Ra, $^{220}$Rn, and $^{216}$Po $\alpha$ decays. The two quenching factors (Sec.~\ref{sec:q-factors}) are simulated and independently fit to data to determine the activity of the $^{228}$Th-group. The fitted results are compared to the time-correlated \alal measurements in Table~\ref{tab:th228-group-activities}. The discrepancies are not well understood. It's possible an $\alpha$ decay isotope is not accounted for in the background model and has an energy in the region of the $^{228}$Th-group causing the fit to overestimate the $^{228}$Th-group activity. 

\begin{table}[ht] \centering
	\begin{tabular}{ c | c | c } 
		& Fitted MC           & \alal coinc.       \\
		& $^{228}$Th-group    & $^{228}$Th-group   \\
		& ($\mu Bq/kg$)       & ($\mu Bq/kg$)      \\
		\hline 
		C1    & 7.47 $\pm$ 0.84    & 3.34 $\pm$ 0.94    \\
		C2    & 4.56 $\pm$ 0.49    & 3.29 $\pm$ 0.06    \\
		C3    & 1.91 $\pm$ 0.20    & 1.69 $\pm$ 0.04    \\
		C4    & 3.93 $\pm$ 0.21    & 2.25 $\pm$ 0.04    \\
		C6    & 2.10 $\pm$ 0.12    & 0.54 $\pm$ 0.02    \\
		C7    & 2.14 $\pm$ 0.12    & 0.58 $\pm$ 0.02    \\
	\end{tabular}
	\caption{Comparison of $^{228}$Th-group activity measured by fitting MC to data and by \alal coincidence as described in Sec.~\ref{sec:aa-events}.}
	\label{tab:th228-group-activities}
\end{table}

\subsubsection{PTFE Surface \texorpdfstring{$^{210}$Po}{210Po}} \label{sec:teflon-surf-po210}
Different $\alpha$ source locations and their resulting spectral shapes were studied with Geant4 simulations. From these studies, it was determined that the low-energy tail ($<$\,2.2\,MeV$_{ee}$) of the $\alpha$ distribution (Fig.~\ref{fig:just-data}) is caused by $^{210}$Po on the surface of the PTFE reflector that surrounds NaI crystals. Furthermore, the depth profile of the $^{210}$Po in the PTFE correlates to the slope of the low energy tail of the $\alpha$ distribution. A shallower depth profile causes a steeper low-energy tail. The contamination depth profiles differ among crystals and range from 2--5\,$\mu$m. The fitted results are summarized in Table~\ref{tab:teflon-surface-activities}. 

Using MC, we investigated the NaI surface $^{210}$Pb at various depth profiles. We also investigated the effects of a scintillation dead layer~\cite{Yang:2014} extending to various depths in combination with NaI surface $^{210}$Pb. These scenarios did not describe the low-energy tail observed in the data. 

Measurements of internal $^{210}$Pb and PTFE surface $^{210}$Pb half-lives were made. The measured half-life of internal $^{210}$Pb was consistent with the known value of 22.3\,yr while the measured half-life of PTFE surface $^{210}$Pb was a much longer 33.8\,$\pm$\,8.0\,yr. One possible explanation is that the surface is continuously contaminated by something radioactive from outside ($^{222}$Rn for example) which artificially increases the observed half-life of PTFE surface~$^{210}$Pb.

\begin{table}[ht] \centering
	\begin{tabular}{ c | c } 
		& PTFE Surf.         \\
		& $^{210}$Po         \\
		& ($\mu Bq/cm^{2}$)  \\
		\hline 
		C1    & 0.77 $\pm$ 0.11    \\
		C2    & 1.25 $\pm$ 0.13    \\
		C3    & 1.10 $\pm$ 0.16    \\
		C4    & 1.08 $\pm$ 0.12    \\
		C6    & 1.61 $\pm$ 0.20    \\
		C7    & 1.14 $\pm$ 0.13    \\
	\end{tabular}
	\caption{PTFE surface $^{210}$Po activity measured by fitting MC to data.}
	\label{tab:teflon-surface-activities}
\end{table}

\section{Conclusion}\label{sec:conc}

%Two quenching factors are required to describe the features in COSINE-100 $\alpha$ data.
Two quenching factors for alpha particles are required to describe the features in COSINE-100 data.
The relative contribution of $Q_{1}$ and $Q_{2}$ is not consistent among fitted isotopes.
Normal-inverse Gaussian PDFs are applied to the simulation to approximate the spectral shapes observed in $\alpha$ data. 
The two quenching factors is hypothesized by the spatial dependence of Tl concentration within the crystal effecting the quenching factor. The asymmetric spectral shape could be due to non-uniform transitions between the two quenching factors. 
More investigation is needed to understand these phenomena.

The COSINE-100 $\alpha$ spectra are well described by dominant components of internal $^{210}$Po, internal $^{228}$Th-group, and PTFE surface $^{210}$Po. NaI surface $^{210}$Po and PTFE bulk $^{210}$Po activities are determined from the \bega channel as described in Ref.~\cite{COSINE-100:2018tfl} because their spectral features are more prominent in the \bega channel. The $\alpha$ isotope activities were measured in this work by fitting Geant4 MC to data. The dominant $^{210}$Po fitted $\alpha$ rates are consistent ($<$\,1.1\,$\sigma$) with the total $\alpha$ rates (Table~\ref{tab:internal-po210-activities}), as measured during commissioning of the crystals.

Furthermore, the half-life of $^{216}$Po has been measured to be $143.4\,\pm\,1.2\,ms$ which is consistent with and more precise than most current measurements.

%There are some peculiarities of the $\alpha$ data. Two quenching factors are required to describe the dominant $\alpha$ peaks (presumably from $^{210}$Po) in data which have a combined $\alpha$ rate consistent with the previously measured internal bulk $^{210}$Pb rate~\cite{COSINE-100:2018tfl}. Furthermore, the spectral shapes of the peaks are highly asymmetric. This phenomena is also observed by the ANAIS experiment~\cite{Amare:2016rbf}, which uses NaI(Tl) crystals produced by Alpha Spectra Inc., the same vendor that produced the crystals for COSINE-100. 

%We explored alternative explanations for the double-peak structure in $\alpha$ data. Using MC, we investigated NaI surface $^{210}$Pb at various depths with a uniform concentration or exponentially weighted concentration. We investigated the effects of a scintillation dead layer~\cite{Yang:2014} extending to various depths in combination with NaI surface $^{210}$Pb. None of these scenarios sufficiently describe the observed $\alpha$ data.

%% Acknowledgments
\section*{Acknowledgments}

%We thank the Korea Hydro and Nuclear Power (KHNP) Company for providing underground laboratory space at Yangyang and the IBS Research Solution Center (RSC) for providing high performance computing resources. 
%This work is supported by:  the Institute for Basic Science (IBS) under project code IBS-R016-A1, NRF-2021R1A2C3010989 and NRF-2021R1A2C1013761, Republic of Korea;
%NSF Grants No. PHY-1913742, DGE-1122492, WIPAC, the Wisconsin Alumni Research Foundation, United States; 
%STFC Grant ST/N000277/1 and ST/K001337/1, United Kingdom;
%Grant No. 2021/06743-1 and 2022/12002-7 FAPESP, CAPES Finance Code 001, CNPq 131152/2020-3, Brazil.

%% new one 2023-09-05
We thank the Korea Hydro and Nuclear Power (KHNP) Company for providing underground laboratory space at Yangyang and the IBS Research Solution Center (RSC) for providing high-performance computing resources. 
This work is supported by:  the Institute for Basic Science (IBS) under project code IBS-R016-A1, NRF-2019R1C1C1005073, NRF-2021R1A2C3010989, NRF-2021R1A2C1013761, NFEC-2019R1A6C1010027, and NRF-2021R1I1A3041453, Republic of Korea;
NSF Grants No. PHY-1913742, DGE-1122492, WIPAC, the Wisconsin Alumni Research Foundation, United States; 
STFC Grant ST/N000277/1 and ST/K001337/1, United Kingdom;
Grant No. 2021/06743-1 and 2022/12002-7 FAPESP, CAPES Finance Code 001, CNPq 131152/2020-3 and 303122/2020-0, Brazil.

%% Appendix
%\appendix
%\section{}
%\label{}

%% Bibliography
\bibliographystyle{elsarticle-num} 
\bibliography{biblio.bib}

\end{document}